\documentclass[aps,prl,reprint,showpacs,twocolumn,superscriptaddress]{revtex4-1}

\usepackage[dvips]{graphicx}
\usepackage{dcolumn}
\usepackage{bm}
\usepackage{xspace}
\usepackage{multirow}
\usepackage{natbib}
\usepackage{color}

\begin{document}

\preprint{}
\title{Hole-Like Fermi Surface in the Overdoped Non-Superconducting Bi$_{1.8}$Pb$_{0.4}$Sr$_2$CuO$_{6+\delta}$}
\author{T. Valla}
\email{valla@bnl.gov}
\affiliation{Condensed Matter Physics and Materials Science Department, Brookhaven National Lab, Upton, New York 11973, USA\\}
\author{P. Pervan}
\affiliation{Institute of Physics, Bijeni\v{c}ka 46, HR-10000 Zagreb, Croatia\\}
\affiliation{Faculty of Science, Department of Physics, University of Split, Ru\-{d}era Bo\v{s}kovi\'{c}a 33, HR-21000 Split, Croatia\\}
\author{I. Pletikosi\'{c}}
\affiliation{Condensed Matter Physics and Materials Science Department, Brookhaven National Lab, Upton, New York 11973, USA\\}
\affiliation{Department of Physics, Princeton University, Princeton, NJ 08544, USA}
\author{I. K. Drozdov}
\author{Asish K. Kundu}
\author{Zebin Wu}
\author{G. D. Gu}
\affiliation{Condensed Matter Physics and Materials Science Department, Brookhaven National Lab, Upton, New York 11973, USA\\}

\date{\today}

\begin{abstract}
In high-temperature cuprate superconductors, the anti-ferromagnetic spin fluctuations are thought to have a very important role in naturally producing an attractive interaction between the electrons in the $d$-wave channel. The connection between superconductivity and spin fluctuations is expected to be especially consequential at the overdoped end point of the superconducting dome. In some materials, that point seems to coincide with a Lifshitz transition, where the Fermi surface changes from the hole-like centered at ($\pi, \pi$) to the electron-like, centered at the $\Gamma$ point causing a loss of large momentum anti-ferromagnetic fluctuations. Here, we study the doping dependence of the electronic structure of Bi$_{1.8}$Pb$_{0.4}$Sr$_2$CuO$_{6+\delta}$ in angle-resolved photoemission and find that the superconductivity vanishes at lower doping than at which the Lifshitz transition occurs. This requires a more detailed re-examination of a spin-fluctuation scenario.

\end{abstract}
\vspace{1.0cm}


\maketitle

\section{Introduction}
Even before the discovery of high temperature superconductors (HTSC), it was realized that antiferromagnetic (AF) fluctuations give rise to a $d$-wave pairing from short ranged repulsion of electrons on a lattice \cite{Hirsch1985,Scalapino1986}. Additionally, for two-dimensional materials with saddle points in the electronic structure, the Van Hove singularities enhance the pairing, if the Fermi level is near the saddle point and the Fermi surface (FS) is not nested \cite{Hirsch1986}. Furthermore, depending on the FS shape, the spin susceptibility will transform from an anti-ferromagnetic, peaked at $q=(\pi, \pi)$ to a ferromagnetic, centered at $q=(0, 0)$ at a filling corresponding to the Van Hove singularity (VHS). This has been considered as the main reason for the loss of pairing in the $d$-wave channel and appearance of ferromagnetic fluctuations \cite{Kopp2007,Kurashima2018} in the strongly hole-overdoped cuprates when the FS undergoes a Lifshitz transition and becomes electron-like, centered at the zone center \cite{Maier2020,Maier2020a}. Similar tendency towards the ferromagnetism at the overdoped end of superconducting dome was also very recently observed in the $n$-type cuprates \cite{Sarkar2020}. The numerical studies of hole-doped cuprates suggest that the pseudogap too is tightly related to the FS topology, existing only for the hole-like FS, with the Lifshitz transition effectively representing its upper phase boundary \cite{Wu2018}. 

\begin{figure*}[t]
\begin{center}
\includegraphics[width=14cm]{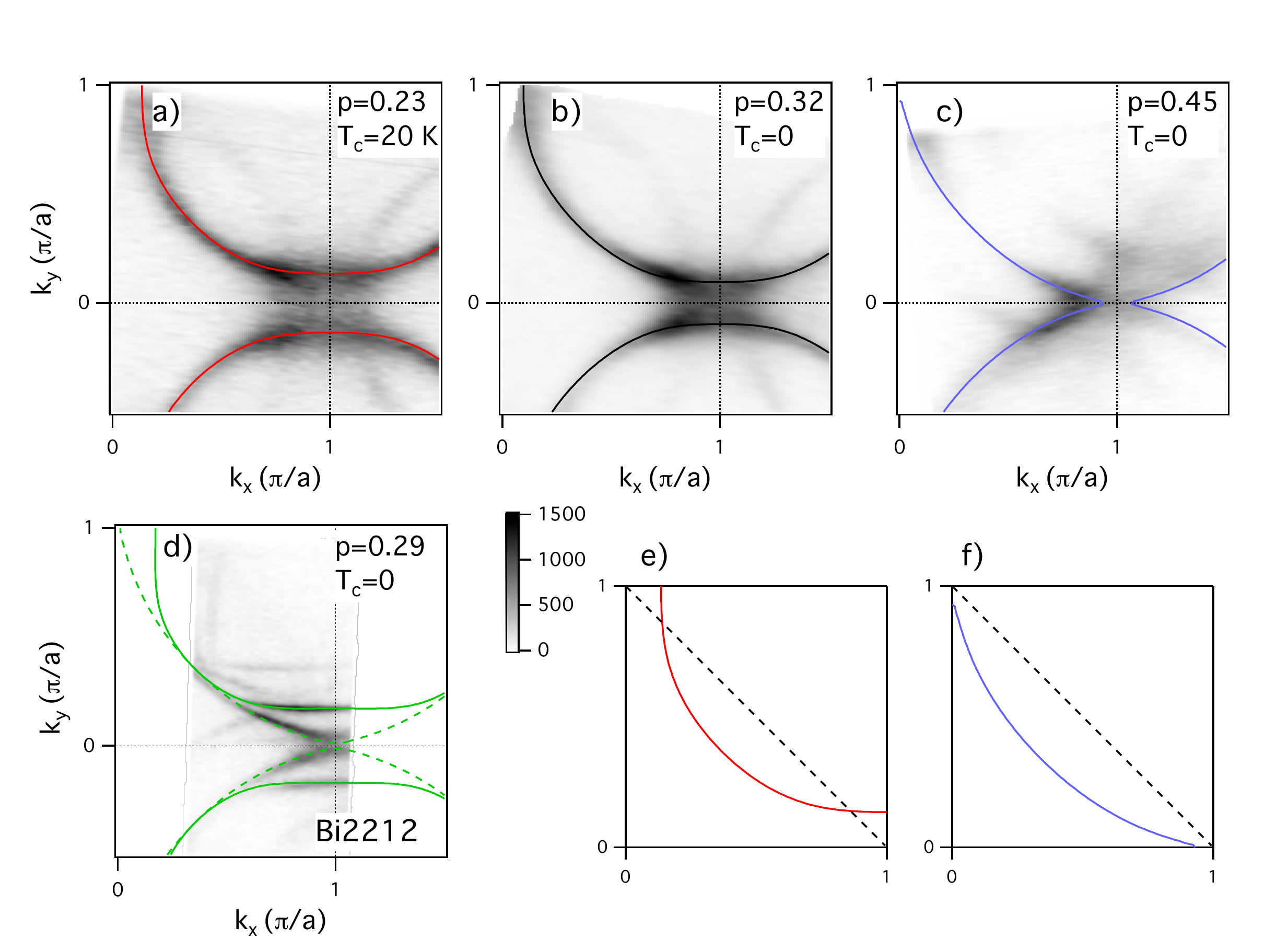}
\caption{Evolution of the Fermi surface of Bi$_{1.8}$Pb$_{0.4}$Sr$_2$CuO$_{6+\delta}$ with doping. (a) Fermi surface of a vacuum annealed bulk sample with the transition temperature $T_c=20$ K. (b) Fermi surface of as grown, non-superconducting sample ($T_c=0$). (c) Fermi surface of as-grown sample after annealing in ozone. (d) Fermi surface of overdoped, non-superconducting Bi2212 ($T_c=0$), for comparison. (e) and (f) illustrate the cases of hole- and electron-like FSs. Spectral intensity in (a-d), represented by the gray-scale contours, is integrated within $\pm3$ meV around the Fermi level. Solid lines represent the states obtained from tight-binding approximation that best fit the experimental data. 
The area enclosed by the TB lines that best represent the experimental data is calculated and used for the determination of the doping parameter $p$. All the maps were recorded at 22 K (normal state).
}
\label{Fig1}
\end{center}
\end{figure*}
%

Experimentally, the coincidental loss of superconductivity and the Lifshitz transition have been observed in some materials at the overdoped end of the superconducting dome, reviving the interest in this matter \cite{Kondo2004,Drozdov2018,Ding2019a,Valla2020,Maier2020}. In the case of the single Cu-O plane cuprates, the coincidence could be considered as a clear-cut evidence that the spin-fermion model is at play at the overdoped endpoint of the doping range \cite{Kondo2004,Ding2019a}. However, the recently established connection between the loss of superconductivity and the Lifshitz transition that was observed in the bilayer system, Bi$_2$Sr$_2$CaCu$_2$O$_{8+\delta}$ (Bi2212) raises more questions \cite{Drozdov2018,Valla2020}. In Bi2212, the Lifshitz transition affects only the anti-bonding FS, but not the bonding one. 
Another cuprate that does not follow this Lifshitz driven scenario is Tl$_2$Ba$_2$CuO$_{6+\delta}$ (Tl2201). The Tl2201 is a clean system showing quantum oscillations and a large FS in the overdoped regime which undergoes the Lifshitz transition at much higher doping than at which the superconductivity disappears \cite{Rourke2010,Putzke2019}. Also, in the case of La$_{2-x}$Sr$_x$CuO$_4$ and La$_{2-x}$Ba$_x$CuO$_4$ the two transitions, although close, are not simultaneous, indicating that this correlation might not be universal for all cuprates \cite{Ino2002,Yoshida2006,Valla2006,Valla2007,Valla2012}. 

These discrepancies call for a reexamination of the spin-fluctuation scenario in which the Lifshitz transition triggers the loss of $d$-wave superconductivity. Here, we study the electronic structure of Bi$_{1.8}$Pb$_{0.4}$Sr$_2$CuO$_{6+\delta}$ (Bi2201), a single layer cousin of Bi2212, in angle-resolved photoemission spectroscopy (ARPES) and find that the superconductivity in this system is lost sooner than the doping at which the Lifshitz transition occurs. This would suggest that the picture in which the topology of the FS alone is the driving force for the loss of superconductivity in the overdoped cuprates is too simplistic and requires other ingredients.

\section{Methods}
The experiments were done in an experimental facility that integrates oxide-MBE with ARPES and SI-STM capabilities within a common vacuum system \cite{Kim2018a,Drozdov2018,Valla2020}. The starting samples were synthesized by the traveling floating zone method and the superconducting transition temperature ($T_c$) was measured with a SQUID down to 1.8 K. The \textit{as-grown} Bi2201 samples were not superconducting and they required vacuum annealing to achieve superconductivity with the maximal $T_c\approx20-28$ K. The \textit{as-grown} and vacuum annealed samples were clamped to the sample holder, cleaved with Kapton tape and studied by ARPES. After the initial measurements, some samples were additionally annealed in vacuum or in ozone, to reduce or increase the hole-doping, respectively. 
The ARPES experiments were carried out on a Scienta SES-R4000 electron spectrometer with the monochromatized HeI (21.22 eV) radiation (VUV-5k). The total instrumental energy resolution was $\sim$ 5 meV. Angular resolution was better than $\sim 0.15^{\circ}$ and $0.4^{\circ}$ along and perpendicular to the slit of the analyzer, respectively. 
\section{Results and Discussion}
Figure \ref{Fig1} shows the photoemission intensity from the narrow energy window around the Fermi level ($\pm$3 meV), representing the Fermi surfaces of the \textit{as-grown} Bi2201 sample, panel (b), the vacuum annealed Bi2201 (a) and the \textit{as-grown} Bi2201 sample after annealing in ozone (c). For comparison, the ozone annealed, non-superconducting Bi2212 sample is also shown in panel (d). The panels (e) and (f) represent the FS that allows the $d$-wave pairing in the antiferromagnetic spin-fluctuation scenario and the one that does not, respectively \cite{Maier2020,Maier2020a}.
The tight binding (TB) contours, representing the best fits to the experimental FSs are also shown \cite{Markiewicz2005}. The bare in-plane band structure is approximated by the tight-binding formula:

$E_{\pm}(k) =\mu - 2t (\cos k_x + \cos k_y) + 4t' \cos k_x \cos k_y - 2t'' (\cos 2k_x + \cos 2k_y) \pm t_{\perp} (\cos k_x - \cos k_y)^2 /4$

For Bi2212, $t_{\perp}\neq0$ and $\pm$ is for anti-bonding (bonding) state; $\mu$ is chemical potential. The hopping parameters that best describe the FSs of the selected measured samples are given in Table \ref{Tab1}.

\begin{figure}
\begin{center}
\includegraphics[width=7cm]{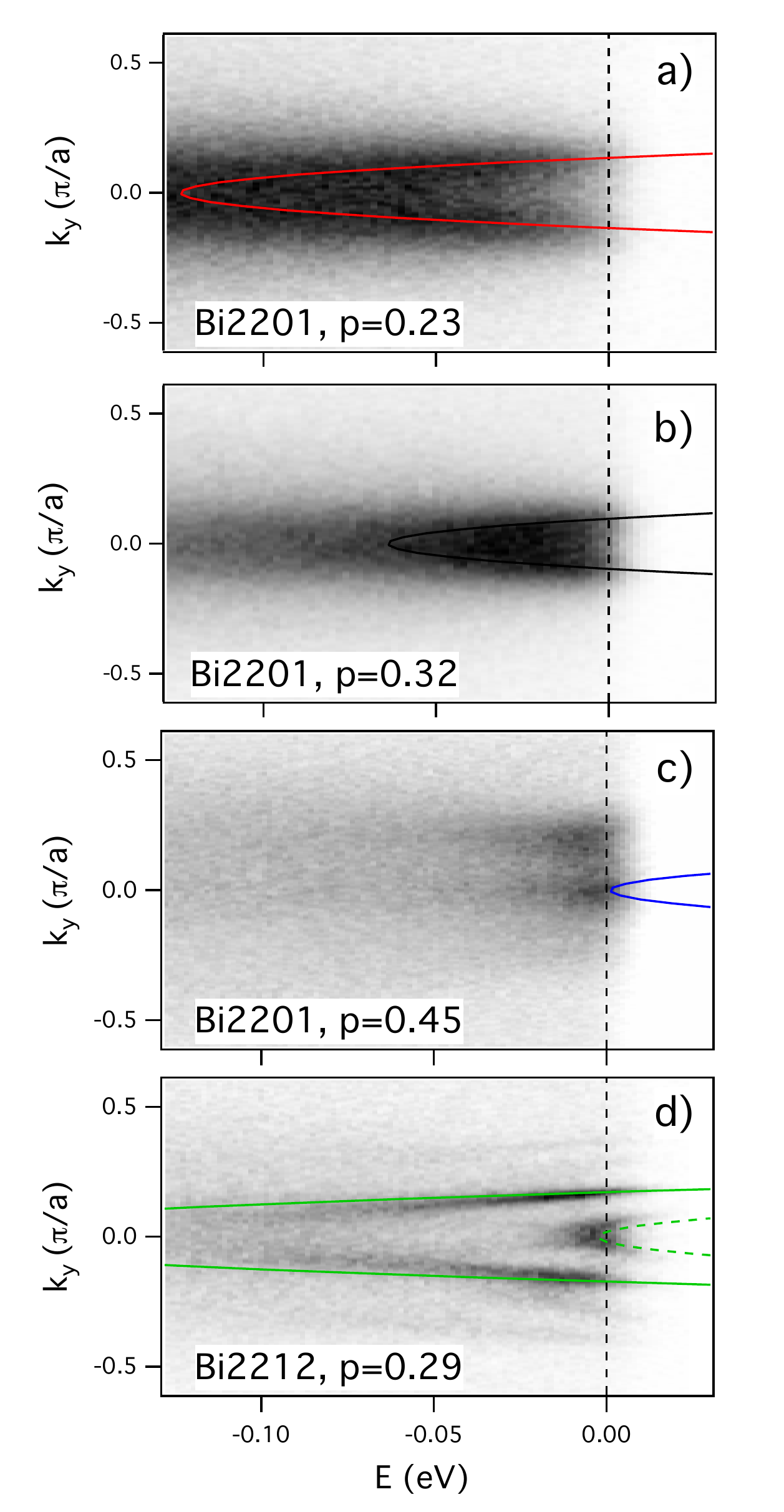}
\caption{Van Hove singularity in Bi$_{1.8}$Pb$_{0.4}$Sr$_2$CuO$_{6+\delta}$. (a-d) Dispersion at $k_x=\pi/a$ along the $Y-M-X$ line corresponding to the Bi2201 and Bi2212 samples from Fig.\ref{Fig1}. The TB dispersions used to fit the Fermi surfaces in Fig.\ref{Fig1} are also shown, following the same color scheme as in Fig.\ref{Fig1}.
}
\label{Fig2}
\end{center}
\end{figure}

The Bi2201 FS consists of a single sheet, while the Bi2212 has two Fermi sheets, the bonding (marked by the solid TB contour) and the antibonding one (dashed TB contour). The \lq\lq{}shadow\rq\rq{} FS, shifted by $(\pi, \pi)$ relative to the intrinsic one is also visible. In addition, in the Bi2212 and in the ozone annealed Bi2201, the supermodulation replicas are also visible. Both the \lq\lq{}shadow\rq\rq{} and the supermodulation replicas are structural in origin and were shown to only minimally affect the electrons in the Cu-O planes \cite{Norman1995,Ding1996b,Mans2006,Valla2019,Gao2020}. The re-appearance of the supermodulation replicas in the ozone annealed Bi2201 is probably related to the segregation of Pb-O and Bi-O in the surface layer. If recleaved, the surface of ozone annealed crystal does not show the supermodulation.
For each sample, the number of doped carriers is obtained directly from the Luttinger count of the area enclosed  by the Fermi contour, $p=2A_{FS}-1$. For the Bi2212, both the bonding and the antibonding states are counted, $A_{FS}=(A_{-}+A_{+})/2$, originating from the two Cu-O planes per unit cell. 

From Fig. \ref{Fig1}(b) it is clear that the Bi2201 FS remains hole-like, centered at $(\pi, \pi)$ at $p=0.32$, even when the superconductivity is already lost. We were able to induce the Lifshitz transition and turn the FS into an electron-like, centered at (0, 0) by pushing additional holes into the Cu-O planes by ozone annealing. This, however, only happens beyond $p\approx0.43$, far away from the doping at which superconductivity is lost. It would appear then that in Bi2201 the two transitions are not bound and that a simple picture where the Lifshitz transition and the loss of $d$-wave pairing occur simultaneously, is not entirely correct. 
This is similar to the Tl2201 case where the superconductivity ceases to exist at $p=0.31$, while the Lifshitz transition is projected to occur at $p\approx0.5$ \cite{Rourke2010,Putzke2019}. 

The Lifshitz transition related to the filling of VHS at the antinodal points of the Brillouin zone is further evidenced in Fig. \ref{Fig2}. The VHS is clearly below the Fermi level for the two less doped samples $(p=0.23$ and 0.32) and only moves slightly above the Fermi level for the ozone-annealed sample ($p=0.45$). For the latter, the state visible at $k_y\approx0.2\pi/a$ is a supermodulation replica of the intrinsic electronic structure, as previously noted (see Fig. \ref{Fig1}(c)).

\begin{table}
\caption{\label{tab}Tight-binding parameters for Bi2201 and Bi2212 samples from Fig.\ref{Fig1}}
\begin{ruledtabular}
\begin{tabular}{lccccc}
Sample & $\mu$ (eV)& $t$ (eV)& $t'$ (eV)& $t''$ (eV)& $t_{\perp}$ (eV)\\
\tableline
Bi2201(a) & 0.42 & 0.36 & 0.1 & 0.036 & 0\\
Bi2201(b) &0.49 & 0.36 & 0.1 & 0.036 & 0 \\
Bi2201(c) & 0.545 & 0.37 & 0.1 & 0.036 & 0\\
Bi2212(d) & 0.467 & 0.36 & 0.108 & 0.036 & 0.108\\
\end{tabular}
\end{ruledtabular}
\label{Tab1}
\end{table}

It is interesting that in the Bi2212, the Lifshitz transition in one of the Fermi sheets (antibonding) coincides with the loss of superconductivity, while the other Fermi sheet (bonding) remains hole-like (Fig. \ref{Fig1}(d) and \ref{Fig2}(d)). Thus, at the end of superconducting dome in Bi2212, we have a situation illustrated in both panels (e) and (f) of Fig. \ref{Fig1} simultaneously. Assuming the simple spin-fluctuation picture were correct, this would imply that the antibonding state is somehow more important for superconductivity than the bonding state. 

\begin{figure}[htbp]
\begin{center}
\includegraphics[width=8.5cm]{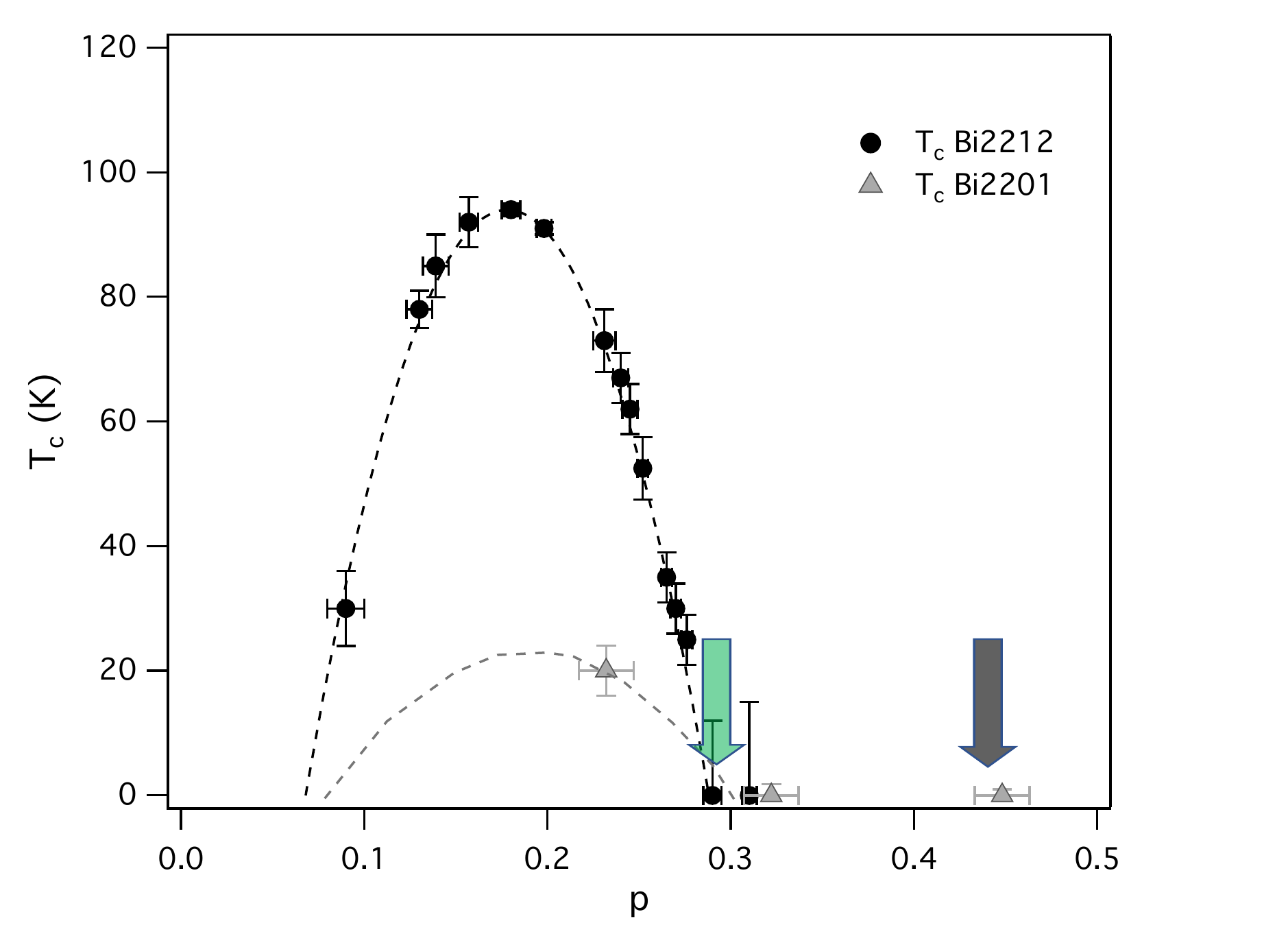}
\caption{Phase diagram of Bi$_{1.8}$Pb$_{0.4}$Sr$_2$CuO$_{6+\delta}$ and Bi$_2$Sr$_2$CaCu$_2$O$_{8+\delta}$. Superconducting transition temperature, $T_c$, for 2212 (black circles) and 2201 (gray triangles) plotted versus experimentally determined doping, $p$. The experimentally determined Bi2212 superconducting dome \cite{Drozdov2018} is represented by black dashed line. The putative dome for 2201 is represented by gray dashed line. The Lifshitz transition in Bi2201 is marked by the gray arrow and the one in the antibonding FS of Bi2212 is marked by the green arrow. 
}
\label{Fig3}
\end{center}
\end{figure}

The main points of this study are summarized in Fig. \ref{Fig3}. We show the doping phase diagram with the superconducting phases for both Bi2212 from Drozdov \textit{et al} \cite{Drozdov2018} and the Bi2201 from this study. We also indicate the position of the Lifshitz transition where the hole like FS turns into the electron-like. In Bi2212 the transition is achievable only in the antibonding Fermi sheet (indicated by the green arrow) and coincides with the end of the superconducting dome. The transition in Bi2201 (gray arrow) occurs far from the superconducting zone boundary. 

We note that our results on Bi2201 are in contrast with previous ARPES studies \cite{Kondo2004,Ding2019a}, probably due to differences in chemical composition of the samples. Kondo \textit{et al} \cite{Kondo2004} report the superconducting phase up to $p\sim0.4$ in Bi2201 co-doped with both Pb and La. In Ding \textit{et al} \cite{Ding2019a} the sample composition is similar to ours and the superconducting phase does not extend that far. Both studies, however, suggest much closer correlation between the loss of superconductivity and the Lifshitz transition. We emphasize  that our \textit{as grown} Bi2201 samples were not superconducting down to 1.8 K, but were very stable both in the ambient conditions and in UHV. They also gave very consistent ARPES results, with the hole-like FS and $p=0.32\pm0.015$ at every cleave ($>10$ cleaves). Therefore, already our \textit{as grown} sample makes the strong case for the title claim of this study.

With the observed irregularities in the coincidence of the Lifshitz transition and the loss of $d$-wave superconductivity in Bi2201 and in different families of cuprate superconductors, a simple picture relating the FS topology with the presence or absence of superconductivity requires a more detailed re-examination. One problem could be that a certain level of three-dimensional (3D) character in the electronic structure ($k_z$ warping) might spread the VHS and place its apparent position in ARPES at the point that depends on the experimental parameters, i.e. the used photon energy. This might be particularly important in La$_{2-x}$Sr$_x$CuO$_4$ that shows a higher level of 3D character \cite{Nakamura1993,Horio2018}, unlike Bi2201 and Tl2201 that are known to be extremely two-dimensional \cite{Manako1991,Ono2003}. 
Another reason for the two transitions occurring non-simultaneously could be that the $T_c$ is reduced much faster than the AF spin-fluctuations due to some other effect. A plausible reason would be a large pair-breaking scattering of some sort that would push the system into the dirty limit and terminate superconductivity before the pairing interactions are gone \cite{Rourke2010,Lee-Hone2017,Pelc2019,Lee-Hone2020,Maier2020a}. A direct comparison between Bi2201 and Bi2212 in Fig. \ref{Fig1} and \ref{Fig2} indeed points to the remarkably broader electronic states in Bi2201 for similar doping levels. That clearly indicates a significantly higher level of scattering in the Bi2201 system that could explain a large mismatch between the Lifshitz transition and the loss of superconductivity in that system. However, the Tl2201 case would argue against this scenario, as this material is clean enough to display quantum oscillations in transport and yet, it shows an even larger mismatch \cite{Rourke2010,Putzke2019,Lee-Hone2020}.

There is also a possibility that the curvature of the FS in the antinodal region might play a significant role in cuprate superconductivity. The straight antinodal segments would normally  favor nesting and a formation of density waves. However, in the highly overdoped regime near the end point of the superconducting dome, the density waves can be discarded, as evidenced by the lack of any gaps or reconstructions that they should produce. These segments might be ineffective in forming mobile Cooper pairs because the group velocity of these segments lacks the component that should connect them from $\mathbf{k_F}$ to $-\mathbf{k_F}$ into singlet pairs. The Bi2212  (Fig. \ref{Fig1}(d)) seems to be a good example of a large discrepancy in curvature of antinodal segments in bonding and antibonding Fermi sheets and an indication that this scenario might be at play in cuprates.

Finally, we speculate that the apparent coincidence of the onset of the ferromagnetism and the demise of superconductivity in the $n$-type cuprates \cite{Sarkar2020}, is also dictated by the same fundamental principles that are expected to be at play in the heavily overdoped $p$-type cuprates. In the superconducting electron doped cuprates, the underlying FS is always hole-like, centered at ($\pi, \pi$), but reconstructed at the \lq\lq{}hot spots\rq\rq{} by the AF spin fluctuations into an electron and a hole pocket. However, at the end point of the superconducting phase in $n$-type cuprates, the FS does not suddenly turn into an electron-like one, as in the case of the hole doped cuprates discussed here. What seems to happen is that the ($\pi, \pi$) centered hole FS becomes so small that it is not touching the AF zone boundary anymore. With the lack of AF \lq\lq{}hot spots\rq\rq{} the FS would not be reconstructed anymore and it should turn into a single hole-like one centered at ($\pi, \pi$) \cite{He2019}. The spin susceptibility is then expected to shift from an antiferromagnetic, with the weight centered at ($\pi, \pi$), to a ferromagnetic, peaked at (0,0). This should be correlated with the loss of $d$-wave pairing, similarly to what happens in the hole-doped cuprates.

\section{Summary}
In summary, we have explored the correlation between the loss of superconductivity and the Lifshitz transition occurring in the heavily overdoped side of the phase diagram of the hole doped cuprates. Although the two transitions are not perfectly coincidental in all the cuprates, there seems to be a general trend where they are relatively close but the exact positions in the doping phase diagram may be influenced by other factors such as the amount of elastic scattering and the curvature of the FS. In addition, the energy scales related to electrons and the details of energy and momentum distribution of spin susceptibility might also play a role in fine tuning of doping positions of these two transitions. This implies that the $d$-wave superconductivity in cuprates is very strongly coupled to the AF spin fluctuations that, when present, likely mediate the pairing in these materials.

\begin{acknowledgments}
This work was supported by the US Department of Energy, Office of Basic Energy Sciences, contract no. DE-SC0012704. I.P. was supported by  ARO MURI program, grant W911NF-12-1-0461. 

\end{acknowledgments}


\end{document}